\documentclass[showpacs,prl,twocolumn,aps,superscriptaddress,preprintnumbers,letterpaper]{revtex4}
\usepackage{amsmath,amssymb}
\usepackage{epsfig}
\usepackage{graphicx}
\usepackage{amsmath}
\usepackage{amsfonts}
\usepackage{epstopdf}
\usepackage{color}

\newcommand{\be}{\begin{equation}}
\newcommand{\ee}{\end{equation}}
\newcommand{\bear}{\begin{eqnarray}}
\newcommand{\eear}{\end{eqnarray}}

\newcommand{\ba}{\begin{array}}
\newcommand{\ea}{\end{array}}

\begin{document}

\title{The origin of thermal component in the transverse momentum spectra \\ in high energy hadronic processes}

\author{Alexander A. Bylinkin}
\affiliation{Institute for Theoretical and Experimental Physics, Moscow 117218, Russian Federation}
\author{Dmitri E. Kharzeev}
\affiliation{Department of Physics and Astronomy, Stony Brook University, Stony Brook, New York 11794-3800, USA}
\affiliation{Department of Physics, Brookhaven National Laboratory, Upton, New York 11973-5000, USA}
\author{Andrei A. Rostovtsev}
\affiliation{National Research Nuclear University MEPhI, Moscow 115409, Russian Federation}

%%%%%%%%%
\date{\today}

\begin{abstract}
The transverse momentum spectra of hadrons produced in high energy collisions can be decomposed into two components: the exponential (``thermal") and the power (``hard") ones.  
Recently, the H1 Collaboration has discovered that the relative strength of these two components in Deep Inelastic Scattering depends drastically upon the global structure of the event - namely, the exponential component is absent in the diffractive events characterized by a rapidity gap. We discuss the possible origin of this effect, and speculate that it is linked to confinement. Specifically, we argue that the thermal component is due to the effective event horizon introduced by the confining string, in analogy to the Hawking-Unruh effect.  In diffractive events, the $t$-channel exchange is color-singlet and there is no fragmenting string  -- so the thermal component is absent. The slope of the soft component of the hadron spectrum in this picture is determined by the saturation momentum that drives the deceleration in the color field, and thus the Hawking-Unruh temperature. 
We analyze the data on non-diffractive $pp$ collisions and find that the slope of the thermal component of the hadron spectrum is indeed proportional to the saturation momentum.  
\end{abstract}


\maketitle

\setcounter{footnote}{0}

%\baselineskip 18pt \pagebreak
%\renewcommand{\thepage}{\arabic{page}}
%\tableofcontents
%\pagebreak

\vskip0.2cm

The transverse momentum spectra of hadrons produced in high energy collisions can be accurately described by the sum of power (``hard") and exponential (``soft") components. The hard component is well understood as resulting from the high momentum transfer scattering of quarks and gluons, and their subsequent fragmentation. The ``soft" one  is ubiquitous in high energy collisions and has the appearance of the thermal spectrum -- but its origin remains mysterious to this day. Indeed, while in nuclear collisions one may expect thermalization to take place, it is hard to believe that thermalization can occur in such processes as Deep-Inelastic Scattering (DIS) or $e^+e^-$ annihilation. Moreover, not only the transverse momentum spectra but also the abundances of hadrons in these elementary processes appear approximately thermal \cite{Becattini:2008tx,Becattini:2010sk,Andronic:2008ev}. 
\vskip0.3cm

The universal thermal character of hadron transverse momentum spectra and abundances in all high energy processes can hardly be a coincidence and begs for a theoretical explanation. One attempt to understand it is based on the hypothesis that confinement is associated with an event horizon for colored particles. The quantum effects then produce the thermal spectra of hadrons, similarly to the Hawking evaporation of black holes or Unruh radiation. Indeed, the color string stretching between the colored fragments in a high energy collision contains the longitudinal chromoelectric field. This field deccelerates the colored fragments producing a Rindler event horizon. Quantum fluctuations in the vicinity of the event horizon then result in the thermal production \cite{Kharzeev:2005iz,Kharzeev:2006zm,Castorina:2007eb}. 
\vskip0.3cm

A novel perspective on this phenomenon is offered by the holographic gauge/gravity correspondence, in which high energy collisions lead to the creation of trapped surfaces (with corresponding event horizons) in the bulk AdS space \cite{Lin:2006rf,Hofman:2008ar,Gubser:2008pc,Hatta:2008qx}. In string approach, the inelastic processes are accompanied by deceleration, and thus the thermal emission \cite{Stoffers:2012zw,Basar:2012jb,Shuryak:2013sra}.
\vskip0.3cm

The effective temperature of the hadron spectrum in this picture is proportional to deceleration that is driven by the confining chromoelectric field. The strength of the chromoelectric field at low collision energies is determined by the string tension. At high energies, the quantum evolution effects come into play, increasing the number of gluons in the wave functions of the colliding hadrons; therefore the chromoelectric field becomes stronger. 
\vskip0.3cm

An economic and theoretically consistent way to describe this phenomenon is offered by the parton saturation \cite{Gribov:1984tu}, or color glass condensate \cite{McLerran:1993ni}, picture. In this approach the density of partons in the transverse plane inside hadrons, and thus the strength of the color field after the hadron collision, is parameterized by the saturation momentum $Q_s(s, \eta)$ that depends on the c.m.s. collision energy squared $s$ and 
(pseudo-)rapidity $\eta$. The deceleration $a$ then appears proportional to the value of the saturation momentum, $a \sim Q_s$. The temperature of the radiation from the resulting Rindler event horizon is thus given by \cite{Kharzeev:2005iz}
\be\label{unruh}
T_{th} = c\ \frac{Q_s}{2 \pi} , 
\ee
where $c$ is a constant of order one; in \cite{Kharzeev:2006zm} an estimate $c \simeq 1.2$ was given. 

\vskip0.3cm

\begin{figure}[h]
\includegraphics[width =8cm]{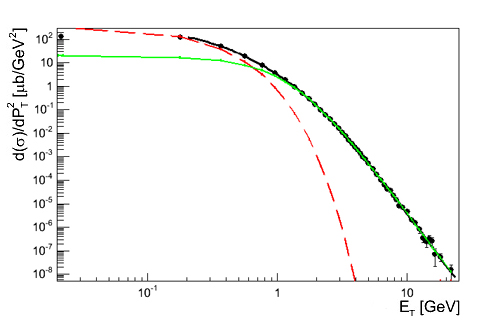}
\caption{\label{fig.0} The charged hadron spectrum measured by the UA1 Collaboration~\cite{UA1} as fitted by the function~(\ref{eq:exppl}): the red (dashed) curve shows the exponential term and the green (solid) one -- the power term.}
\end{figure}

The dependence of the saturation momentum on c.m.s. energy squared $s$ and pseudo-rapidity $\eta$ is given by
\be\label{sat}
Q_s^2 (s; \pm \eta) = Q_s^2( s_0; \eta = 0)\ \left(\frac{s}{s_0}\right)^{\lambda/2}\ \exp(\pm \lambda \eta) ;
\ee
where $\lambda \simeq 0.2 \div 0.3$ is the intercept (see e.g. \cite{Kharzeev:2001gp}). 
 In the saturation scenario, $Q_s$ is the only dimensionful parameter, so the transverse momentum spectra $F(p_T)$ have to scale as a function of dimensionless variable $p_T/Q_s$ \cite{SchaffnerBielich:2001qj,McLerran:2010ex}:
\be
F(p_T) = F(p_T/Q_s) ;
\ee
for massive hadrons of mass $m$, we have to replace $p_T \to m_T = \sqrt{p_T^2 + m^2}$. 
\vskip0.3cm
In ref. \cite{Bylinkin:2012} it has been found that the following parameterization describes well the hadron transverse momentum distribution in hadronic collisions and deep-inelastic scattering:
\be
\label{eq:exppl}
\frac{d\sigma}{p_T d p_T} = A_{therm} \exp {(-m_{T}/T_{th})} +
\frac{A_{hard}}{\left(1+\frac{m_T^2}{T^{2}\cdot n}\right)^n},
\ee
A typical fit to the charged particle spectrum with this function ~(\ref{eq:exppl}) is shown in the Fig.~\ref{fig.0}.

Within the framework described above, the parameter  
$T$ is the saturation momentum, $Q_s = T$, and the effective temperature $T_{th}$ is proportional to $Q_s$ as well, as given by (\ref{unruh}). Therefore, basing on the picture outlined above, we expect the linear relation between $T_{th}$ and $T$. The linear relation $T = (4.26\pm0.15)\cdot T_{th}$ has indeed  been observed in \cite{Bylinkin:2012}.
\vskip0.3cm
Moreover, since the presence of the thermal component signals deceleration in longitudinal color fields, we can now understand a striking experimental observation \cite{Bylinkin:2014fta}: in diffractive events characterized by a rapidity gap, the thermal component in the hadron transverse momentum spectrum is absent. In our present framework, this is a straightforward consequence of the color-singlet $t$-channel exchange that is responsible for diffraction --  in this case there is no fragmenting string  -- and thus no deceleration.
\vskip0.3cm
Let us now check whether the relation between $T_{th}$ and $T$ is indeed linear at different energies and rapidities. 
Since the parameters $T$ and $T_{th}$ according to (\ref{sat}) should depend on both c.m.s. energy and pseudo-rapidity, it is desirable to separate these  dependences. This is possible if one compares the data at different c.m.s. energies at approximately the same pseudorapidity intervals. Hence we look first at ISR~\cite{ISR}, PHENIX~\cite{PHENIX}, ALICE~\cite{ALICE} and UA1~\cite{UA1} data in the most central ($|\eta|<0.8$) pseudo-rapidity region. 
\vskip0.3cm

\begin{figure}[h]
\includegraphics[width =8cm]{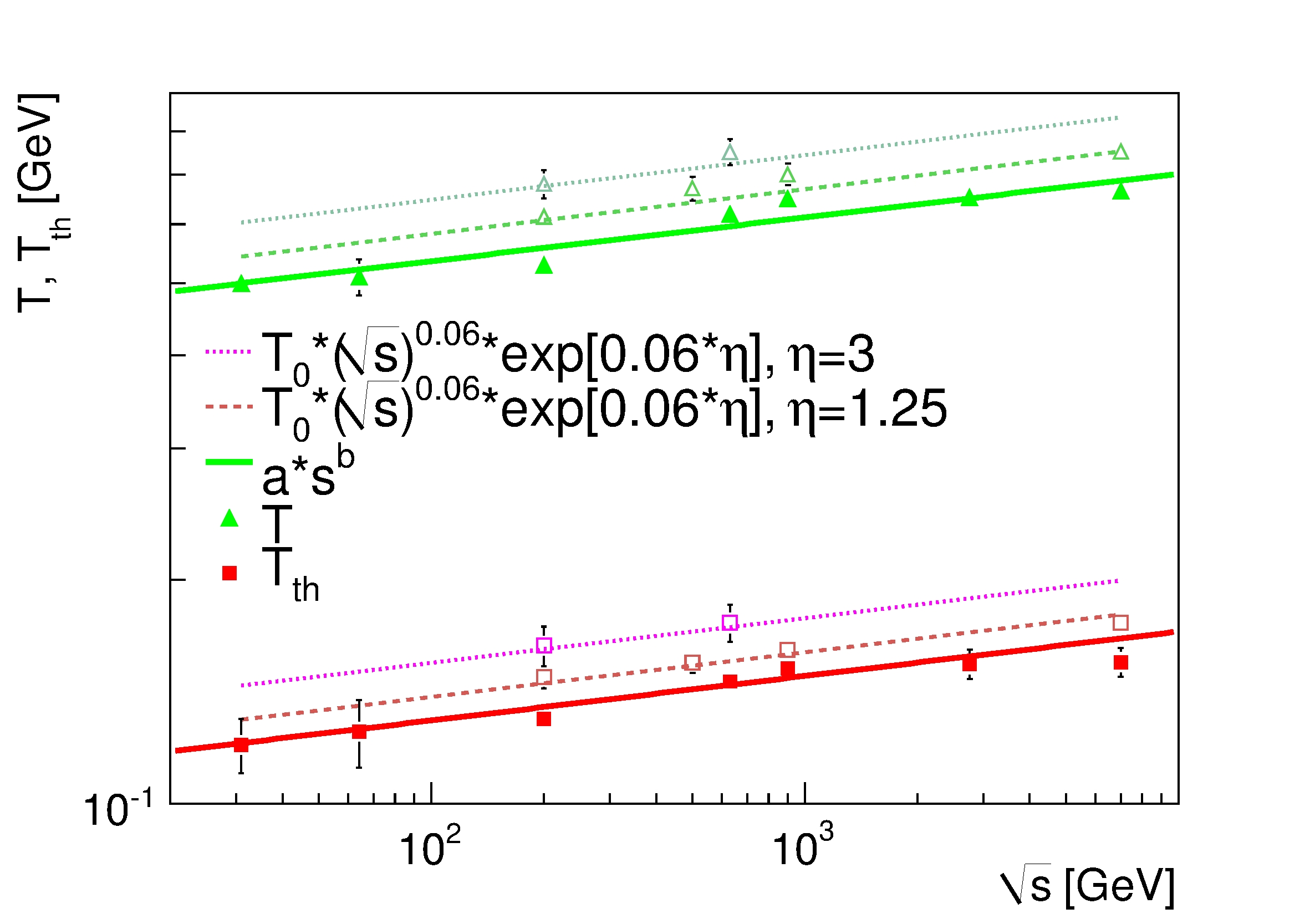}
\caption{\label{fig.1} Variations of the $T$, $T_{th}$ parameters (\ref{eq:exppl}) obtained from the fits to the experimental data~\cite{ISR, PHENIX, UA1, ALICE} (full points) as a function of c.m.s. collision energy $\sqrt{s}$. Solid lines show power-law fits (\ref{sat}). In addition, open points show parameters for the data measured in different pseudorapidity intervals~\cite{UA1,UA11,CMS, BRAHMS} with dashed and dotted lines showing the predictions from (\ref{sat}).}
\end{figure}

\begin{figure}[h]
\includegraphics[width =8cm]{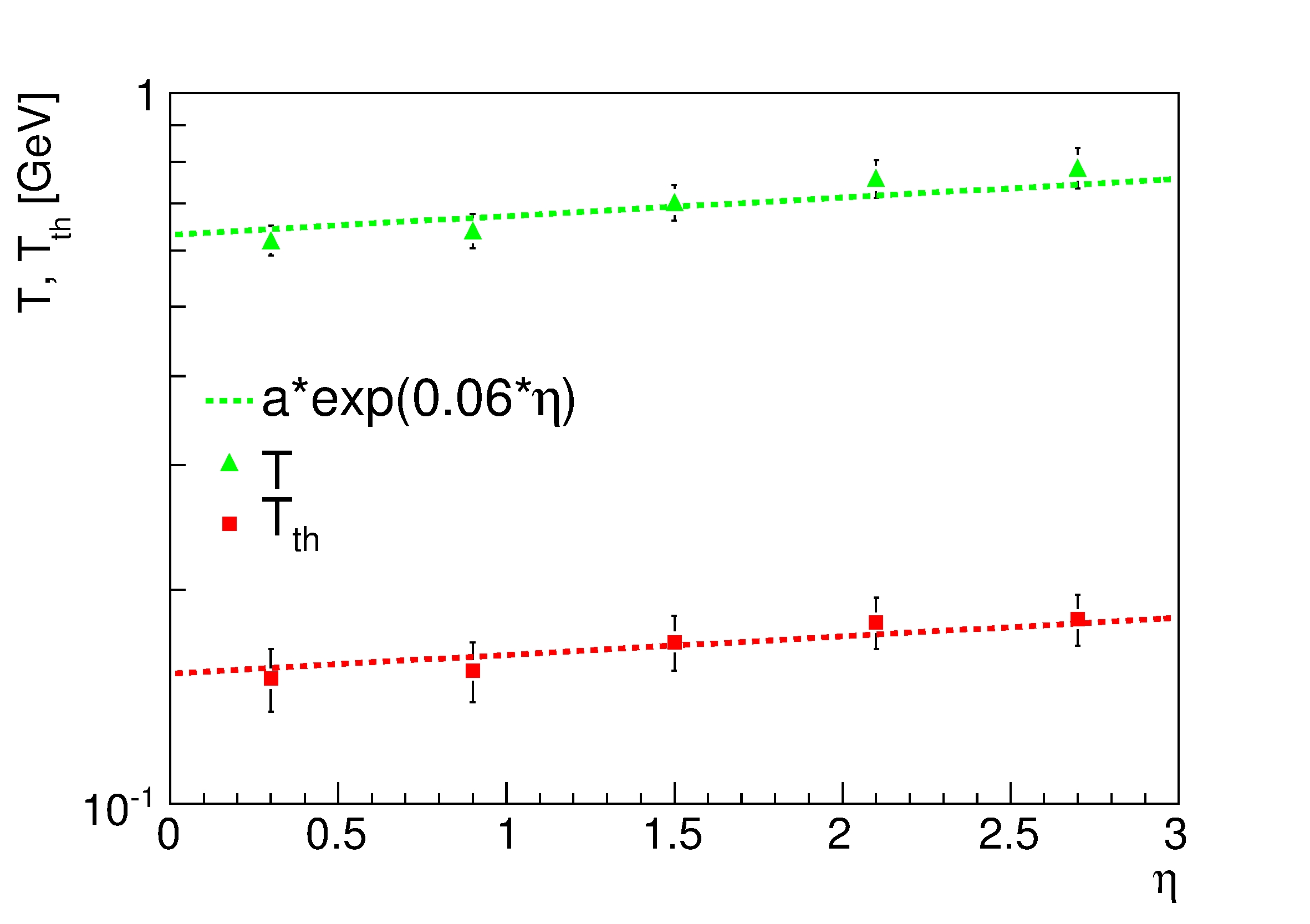}
\caption{\label{fig.2} Variations of the $T$ and $T_{th}$ parameters of (\ref{eq:exppl}) obtained from the fits to the experimental data~\cite{UA1} as function of pseudorapidity. Lines show the dependence predicted by eq.~(\ref{sat}) with $\lambda=0.12$.}
\end{figure}

The figure~\ref{fig.1} shows the values of $T$ and $T_{th}$ resulting from this analysis as a function of c.m.s. collision energy. 
One can describe the energy dependence by the power-law fits (\ref{sat}) shown in figure~\ref{fig.1}:
\be
\label{eq.st}
T =  409 \cdot (\sqrt{s})^{0.06}~MeV,
\ee
\be
\label{eq.ste}
T_{th} = 98\cdot(\sqrt{s})^{0.06}~MeV.
\ee
We find a rather good agreement between the values extracted from the fit (\ref{eq:exppl}) of experimental data and expected on the basis of (\ref{sat}). 
Remarkably, from (\ref{eq.st})-(\ref{eq.ste}) one can again notice the linear relation between $T$ and $T_{th}$ with the proportionality coefficient $4.16\pm0.2$, similar to that observed in~\cite{Bylinkin:2012}.
\vskip0.3cm
The extracted proportionality coefficient $4.16$ is not far from $(2 \pi)/1.2 \simeq 5.23$ predicted in \cite{Kharzeev:2006zm}, so $c$ is indeed of order 1.
\vskip0.3cm
To study the variations of parameters $T$ and $T_{th}$ with pseudorapidity one can use the data from the UA1 experiment~\cite{UA1} which are available as the charged hadron spectra in five pseudorapidity bins, covering the total rapidity interval $|\eta|<3.0$. Figure~\ref{fig.2} shows how the parameters $T$ and  $T_{th}$ vary with pseudorapidity together with the lines describing the exponential dependence predicted from eq.~(\ref{sat}) with $\lambda=0.12$ as obtained from the fits (\ref{eq.st})-(\ref{eq.ste}) to the experimental data.
Though the data measured by the UA1 experiment have been measured only in five pseudorapidity intervals, one can clearly notice the growth of $T$ and $T_{th}$ values, which is also in a good qualitative agreement with the formula (\ref{sat}). Further precise measurements of the double differential charged particle spectra at LHC would allow to confirm the observed behavior. 

In addition, figure~\ref{fig.1} shows UA1~\cite{UA1,UA11}, BRAHMS~\cite{BRAHMS} and CMS~\cite{CMS} data measured under different experimental conditions. In these measurements the pseudo-rapidity interval was much wider than in~\cite{ISR,PHENIX,ALICE}. Therefore, one can compare the parameter values obtained from the fits of these data (open points in figure~\ref{fig.1}) to the values calculated according to (\ref{sat}) with $\lambda=0.12$, $T^0$ and $T_{th}^0$ taken from (\ref{eq.st})-(\ref{eq.ste}) and $\eta$ taken as the mean value of the measured pseudorapidity interval (dashed and pointed lines in figure~\ref{fig.1}. Rather good agreement between these predictions and the experimental data can be observed from figure~\ref{fig.1} further supporting the behavior predicted by eq.~(\ref{sat}).
\vskip0.3cm
We hope that our analysis sheds some new light on the origin of the thermal component in hadron production. The established proportionality of the parameters describing the ``thermal" and ``hard" components of the transverse momentum spectra supports the theoretical picture in which the soft hadron production is a consequence of the quantum evaporation from the event horizon formed by deceleration in longitudinal color fields. The absence of the thermal component in diffractive interactions lend further support to our interpretation. It will be worthwhile to extend this analysis to other high energy processes. Future precise measurements at LHC are needed to further study the proposed picture for hadron production.

\vskip0.3cm

  The work of D.K. was supported in part by the U.S. Department of Energy under Contracts No.
DE-FG-88ER40388 and DE-AC02-98CH10886.

%%%%%%%%%%%%%%%%%%%%%%%%%%%%%%%%%%%%%%%%%%%%%%%%%%%%%%%%%%%%%%%%%%%

\end{document}